\documentclass[aps,pra,twocolumn,showpacs,preprintnumbers,amsmath,amssymb,groupedaddress]{revtex4}
\usepackage{graphicx}

\usepackage{txfonts}
\usepackage{amssymb}

\begin{document}


\title{Vacuum Fluctuation (1): the Same Basis of the Relativity and the Quantum Mechanics}


\author{Xing-Hao Ye}
\email[Electronic address:]{yxhow@163.com}

\affiliation{Department of Physics, Zhejiang University, Hangzhou
310027, China }


\date{\today}

\begin{abstract}
The aim of this paper is to reveal the deep relationship between
matter and vacuum, and to seek for the same physical basis of the
relativity and the quantum mechanics. In doing this, three
postulates of vacuum fluctuation are proposed first, the basic
premises of the relativity and the quantum mechanics including the
velocity limit, the energy-frequency relation and the de Broglie
wavelength expression of any matter particles are deduced then. As
applications, the idea is used to analyze the Compton effect and the
electron-positron annihilation. It is found that the calculation
becomes simple, and the physical meaning gets clear. The simplicity
comes from the power of the three postulates. To illustrate this,
the basic conclusions of the special theory of relativity such as
the relations of mass-velocity, mass-energy, energy-momentum, time
dilation and length contraction are further deduced. In addition,
the significance of the investigation of vacuum fluctuation in the
unification of the physical theories is pointed out.
\end{abstract}


\maketitle


\section{Introduction}
\label{intro} The theory of relativity and the quantum mechanics are
the two bases of the twentieth century physics which has achieved
great success and made strong impact on many other fields. Since the
two theories have been tested widely in practice, there is no doubt
about their correctness. Yet, they are not the ultimate theories. As
we know that, the two theories answered only the question ``how''
rather than ``why''. For example, the theory of relativity pointed
out only that the velocity of light in vacuum is a maximum and a
constant, without giving its reason; and the quantum mechanics
described only the probability of microcosmic particles, but not its
mechanism. Besides, there is only a calculational combination of the
two theories, but not a physical unification. Therefore,
difficulties arise when fundamental problems are considered. Today,
problems such as the interference of a single photon
\cite{Zeilinger2005}, the quantum entanglement and nonlocal
correlation \cite{Ikram2007,Boer2005}, the origin of mass
\cite{Weiglein2004,Wilczek2005}, the creation of particles
\cite{Lamb2007}, the confinement of quarks \cite{Baldo2007}, the
origin and evolution of the universe \cite{Bojowald2005}, the nature
of dark matter and dark energy \cite{Bennett2006,Bennett2005}, the
Hawking radiation of black holes \cite{Gibbons2002,Iso2006}, the
nature of space and time \cite{Bojowald2005,Campo2003}, the relation
between the four fundamental interactions and the unification of the
theory of gravity and the quantum mechanics
\cite{Camelia2000,Hall2002} etc., are all calling for the
investigation of the true foundation of the existing theories.

On the other hand, people have long thought that the most
fundamental problems of physics may all be related to the quantum
vacuum. It is believed that vacuum will be an entrance to the
complete understanding of the physical world and to the ultimate
unification of the physical theories.

One of the fundamental problems is: ``What are the fundamental
elements of the physical world?'' Today, there are too many members
of fundamental particles we have discovered, so they are not truly
fundamental. Some scientists suggest that strings or quantum loops
may be the fundamental ones. But still, there are too many types of
strings or loops in their theories, and we may ask further that:
``What are those strings or loops composed of?'' In principle, the
fundamental elements of the world must be simple in characteristic
and minimal in type, just as the binary numbers ``0'' and ``1'' can
form every type of information in computer science. The ancient
Chinese philosophy and the modern quantum theory may give us
revelation. In Chinese Taoism, Tao produces Yuanqi (original energy)
and Yuanqi is divided into Yin (the dark or negative side) and Yang
(the light or positive side). Yin and Yang are opposite and mutually
complementary, and the balance or harmony of Yin and Yang forms the
essence of everything \cite{Yin-Yang}. In modern quantum theory of
field, particles are regarded as states excited from the vacuum,
that is to say, particles and antiparticles can be excited or
created from the quantum vacuum under certain condition. The idea
was first proposed by Dirac in 1928 and 1930. In his papers, vacuum
was regarded as a ``negative energy sea'', from which electron could
be excited and antielectron could be formed\cite{Dirac1928-1930}.
The predicted antielectron was discovered soon by Anderson in 1932
\cite{Weinberg1995}. After then, many antiparticles were found in
succession. The fact leads to the thought that all the matters may
be originated from vacuum and vacuum may be composed fundamentally
of two opposite elements. It means that vacuum is not just an empty
void, but a special physical existence --- this has been verified to
be true. In 1948, Casimir proposed that the zero point fluctuation
of vacuum can be measured through an attractive effect
\cite{Casimir1948}, which was examined experimentally by Lamoreaux
in 1997 \cite{Lamoreaux1997}. Also, there are other phenomena that
show the non-empty property of the vacuum, among which are the
electron anomalous magnetic moment and Lamb shift \cite{Ahmadi2006},
the light polarization rotation in vacuum in the presence of a
transverse magnetic field \cite{Zavattini2006}, etc.

A second fundamental problem is: ``What is gravitation? Can
gravitation be interpreted on a more fundamental basis and thus be
unified with other fundamental forces?'' Early in 1920, Einstein
\cite{Einstein1920} stated that ``according to the general theory of
relativity space is endowed with physical qualities; in this sense,
therefore, there exists an ether.'' In 1920, Eddington
\cite{Eddington1920} proposed that the space could be regarded as a
special optical medium whose refractive index varies with the
distance to the gravitational matter and that the deflection of
light in a gravitational field could be interpreted as an effect of
refraction in a flat spacetime. This view of gravitational space was
developed later by Wilson \cite{Wilson1921}, Dicke \cite{Dicke1957},
Felice \cite{Felice1971}, Nandi \emph{et
al}.\cite{Nandi1995,Evans1996,EvansNandi1996}. Recently, this way in
treating the gravitational force has been investigated further by
Puthoff \cite{Puthoff2002,Puthoff2005}, Vlokh \cite{Vlokh2004-2006},
etc \cite{Ye2007}. In their view, the ether or space stated above is
just related to the quantum vacuum, and through the study of which,
a true understanding of the gravitation would be acquired.

Besides the above, there are other fundamental problems such as
``What is inertia?'', ``What is mass'', etc., being considered to be
related to the quantum vacuum
\cite{Haisch1994,Haisch1998,Haisch2000,Dobyns2000,Rueda2005}.

So it is reasonable to say that vacuum is a key to the understanding
of nature!

The purpose of this paper is to find out the deep relationship
between matter and vacuum, and to quest for the same physical basis
of the theory of relativity and the quantum mechanics. To reach this
aim, three postulates of vacuum fluctuation will be proposed first,
which are quite simple and natural.

\section{Basic postulates of vacuum fluctuation}
\label{sec:1} \subsection{Postulate 1: Matter is a buildup of vacuum
fluctuational particles}

There are three questions associated with this postulate:

(1) What is a vacuum fluctuational particle?

A vacuum fluctuational particle (it will be called a ``vacuflucton''
bellow for short) is a most fundamental particle excited from the
quantum vacuum. Basically, there are two types of vacufluctons: if
we call one  ``positive vacuflucton'', then the other  ``negative
vacuflucton'' --- each is the counterpart of the other (Fig.1).
Vacufluctons are not fixed. They randomly emerge from and vanish
into the vacuum.
\begin{figure}
\centering
\includegraphics[totalheight=0.6in]{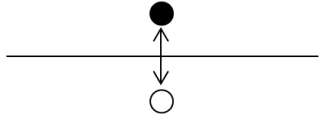}
\caption{\label{fig01} Positive-negative vacufluctons.}
\end{figure}

In Fig.\ 1, the horizontal line represents the vacuum, if
``$\medbullet$'' represents a positive vacuflucton, then ``$\medcirc
$'' a negative vacuflucton. When such a pair of positive-negative
vacufluctons forms, there is an orientational motion of the matter
in space by a mechanism of the vacufluctons vanishing here and
emerging there; whereas a single type of vacufluctons could only
have a random change of places, with no orientational motion formed.

It is necessary to point out that the ``vacuum fluctuation''
discussed here is different to the commonly said ``zero-point
fluctuation of the vacuum''. The properties that there are two
opposite types of vacufluctons (``positive'' and ``negative'') and
that the vacufluctons can randomly ``emerge from and vanish into the
vacuum'' indicate that the ``vacuum fluctuation'' discussed here is
somewhat like the ``vacuum polarization'', ``vacuum excitation'', or
``vacuum vibration''. But still, they are different concepts ---
when we say ``vacuum fluctuational particles'' or ``vacufluctons'',
we are investigating the more microscopic composition of matter and
stressing its ``growth and decline'' or ``eat and flow''
relationship with vacuum.

(2) What is a ``fundamental particle'' composed of?

A commonly said ``fundamental particle'' such as a photon, an
electron and others is composed of a large number of (that is, a
group of) vacufluctons. Because of the random emerging and vanishing
of the vacufluctons themselves, such a composition is quite
complicated. Fig.\ 2 shows a simplified construction of a
``fundamental particle''.
\begin{figure}
\centering
\includegraphics[totalheight=1.0in]{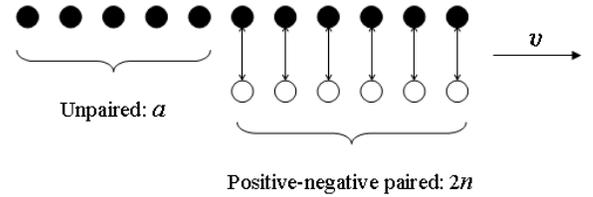}
\caption{\label{fig02} A group of vacufluctons form a ``fundamental
particle''.}
\end{figure}

In Fig.\ 2, the $a$ unpaired  vacufluctons represent the rest
matter. When they are added with $2n$ positive-negative paired
vacufluctons, the whole group of vacufluctons corresponding to the
moving matter then have an orientational motion in space at velocity
$\upsilon$. Hence, the total number of vacufluctons within a matter
particle can be expressed generally as
\begin{equation}
Q=a+2n.
\end{equation}

Since the relativity of motion, the vacufluctons are also relative.
That is, a certain number of vacufluctons observed in a reference
frame may have a different number of vacufluctons in other frames.
But in one frame, the total number of vacufluctons of an isolated
system (containing particles 1, 2, ..., $i$, ...) is invariable,
that is
\begin{equation}
d\sum Q_i=0.
\end{equation}

(3) What is the mass of a particle?

The mass of a particle reflects the quantity of matter. Therefore,
the mass of a particle is directly proportional to the total number
of vacufluctons it contains
\begin{equation}
m=k_1 Q,
\end{equation}
where $k_1$ is a constant.

\subsection{Postulate 2: The orientational motion of a group of vacufluctons forms the momentum of a particle}

Because the orientational motion of a group of vacufluctons relies
on the pairs of positive-negative vacufluctons, the momentum of a
particle is then determined by the possibility of positive-negative
pairing. It is postulated that
\begin{equation}
\vec{p}=m\vec{\upsilon}=k_2 \vec{\sqrt{(a+n)n}},
\end{equation}
where $k_2$ is also a constant. The total possibility of random
pairing of positive-negative vacufluctons is figured out as
$(a+n)n$. For instance, in Fig.\ 3, each negative vacuflucton can be
the partner of $a+n$ positive vacufluctons; therefore the total
number of possible partnership is $(a+n)n$.
\begin{figure}
\centering
\includegraphics[totalheight=1.4in]{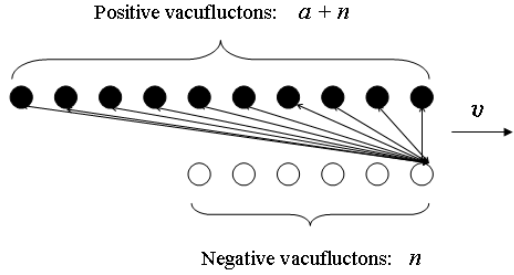}%
\caption{\label{fig03}The random pairing of positive-negative
vacufluctons.}
\end{figure}

Because of the relativity of vacufluctons, the momentum of a
particle is also relative. While in a certain frame of reference,
the total momentum of an isolated system is conserved:
\begin{equation}
d\sum \vec{p}_i=0.
\end{equation}

\subsection{Postulate 3: The frequency of fluctuation is proportional to the number of vacufluctons}

Each fluctuation, that is, each emerging and vanishing of a
vacuflucton, is an interaction or a vibration between matter and
vacuum. Therefore, the frequency of fluctuation of the group of
vacufluctons is directly proportional to the total number of
vacufluctons:
\begin{equation}
\nu =k_3 Q,
\end{equation}
where $k_3$ is another constant.

\section{Deductions of the premises of the relativity and the
quantum mechanics} \label{sec:2}
\subsection{Velocity limit of a moving particle}

Eq. (4) gives $m\upsilon=k_2 \sqrt{(a+n)n}$. Substituting Eqs. (1)
and (3) yields
\begin{equation}
\upsilon=\frac{k_2}{k_1} \frac{\sqrt{(a+n)n}}{a+2n}.
\end{equation}

Through this relation, the condition of velocity limit can be
obtained as $a/n=0$, or $a=0$, or $n=\infty$. Then the maximum
velocity of a moving particle will be
\begin{equation}
\upsilon_{max}=\frac{k_2}{k_1}
\frac{\sqrt{(0+n)n}}{0+2n}=\frac{k_2}{2k_1}.
\end{equation}

It shows that a particle of zero rest mass moves at a maximum
velocity of $k_2/2k_1$, which is a uniform constant no matter which
reference frame is selected. And we know that a photon just
satisfies this property. Therefore the velocity $c$ of a photon in
vacuum is the velocity limit for any moving particle:
\begin{equation}
\upsilon_{max}=\frac{k_2}{2k_1}=c.
\end{equation}

This deduction corresponds to the basic premise of the special
theory of relativity.

\subsection{Energy-frequency relation}

Eqs. (4) and (9) give
\begin{equation}
\vec{p}=m\vec{\upsilon}=2ck_1 \vec{\sqrt{(a+n)n}}.
\end{equation}

Through the definitions of force $\vec{F}=d\vec{p}/dt$, work
$dW=\vec{F}\bullet d\vec{s}$, and kinetic energy $dE_k=dW$, we get
\begin{equation}
dE_k=(d\vec{p})\bullet \vec{\upsilon}=\upsilon\cdot dp.
\end{equation}

Substituting Eqs. (7), (9) and (10) into the above gives
\begin{equation}
dE_k=2c\frac{\sqrt{(a+n)n}}{a+2n} \cdot d[2ck_1 \sqrt{(a+n)n}].
\end{equation}

For a moving particle, $a$ is a fixed number, while $n$ is a
variable. Thus we get $dE_k=c^2d(k_1\cdot 2n)$, or $
dE_k=c^2d[k_1\cdot (a+2n)]$. Associating this with Eqs. (1) and (3)
gives
\begin{equation}
dE_k=c^2dm.
\end{equation}

Then, through the integration
\begin{equation}
E_k=\int^{m}_{m_0} c^2dm,
\end{equation}
we get $E_k=mc^2-m_0 c^2$, or $E=E_0+E_k=m_0 c^2+E_k=m c^2$, where
$m_0$ is the rest mass, $E_0=m_0 c^2$ is the rest energy. So there
is the mass-energy relation
\begin{equation}
E=m c^2.
\end{equation}

Associating Eqs. (3), (6)and (15) gives
\begin{equation}
E=(\frac{k_1}{k_3}c^2)\nu.
\end{equation}

According to the law of blackbody radiation, the constant here is
just the Planck constant $h$:
\begin{equation}
(\frac{k_1}{k_3}c^2)=h.
\end{equation}

Then we have the energy-frequency relation
\begin{equation}
E=h \nu.
\end{equation}

\subsection{de Broglie wavelength expression}

Through Eqs. (11), (13) and (15) we know that the velocity of a
moving particle is $\upsilon =dE/dp$. In the view of wave motion, it
is the group velocity of the vacufluctons of the particle.
Considering
\begin{equation}
\upsilon=\frac{dE}{dp}=\frac{d(h\nu)}{dp}=\frac{d\omega}{d(\frac{2\pi
p}{h})},
\end{equation}
and the general expression of group velocity
\begin{equation}
\upsilon=\frac{d\omega}{dk},
\end{equation}
we have
\begin{equation}
\frac{2\pi p}{h}=k=\frac{2\pi}{\lambda}.
\end{equation}
Thus we get
\begin{equation}
\lambda=\frac{h}{p}.
\end{equation}
It is just the de Broglie wavelength expression.

\

The above two deductions, that is, the relations of energy-frequency
and momentum-wavelength, show the wave-particle duality of any
matter particles, which is the basic premise of the quantum
mechanics. And we find that the two properties, that is, the wave
property ($\nu$, $\lambda$) and the particle property ($E$, $p$) ,
are connected through the Planck constant: $E/\nu=h$, $p\lambda=h$.
As a matter of fact, the Planck constant characterizes the
generality and unity of vacuum fluctuation. We can see this from the
the value of the Planck constant. Eqs. (9) and (17) give the
relation $h=k_2^2/4k_1k_3$. So the Planck constant $h$ is an
integration of the three constants $k_1$, $k_2$ and $k_3$ of the
vacuum fluctuation.

\section{Applications}
\label{sec:3}
\subsection{Compton effect}

First, consider the numbers of vacufluctons before and after the
interaction between the photon and electron (Fig.4). According to
Postulate 1, the total number of vacufluctons of the system is
conserved:
\begin{figure}
\centering
\includegraphics[totalheight=1.8in]{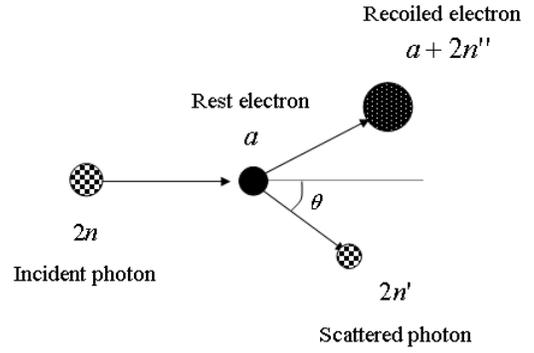}
\caption{\label{fig04} Conservation of matter in Compton effect.}
\end{figure}
\begin{equation}
2n+a=2n'+(a+2n''),
\end{equation}
where $2n$ and $2n'$ are the numbers of vacufluctons of the photon
before and after the interaction respectively, $a$ and $a+2n''$ are
those of the electron. Then we have
\begin{equation}
n=n'+n''.
\end{equation}

Second, consider the momentum of the system before and after the
interaction. According to Eq.(4), the momentum ratio between the
photon before the interaction, the photon and the electron after the
interaction is $n:n':\sqrt{(a+n'')n''}$. Fig.5 shows this relation
of the three momenta. Then, according to Postulate 2, the momentum
of the system is conserved, we have
\begin{figure}
\centering
\includegraphics[totalheight=0.78in]{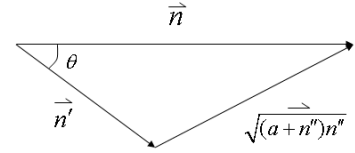}
\caption{\label{fig05} Conservation of momentum in Compton effect.}
\end{figure}
\begin{equation}
\vec{n}=\vec{n}'+\vec{\sqrt{(a+n'')n''}}.
\end{equation}

Associating Eqs.(24) and (25) gives
\begin{equation}
\frac{1}{n'}-\frac{1}{n}=\frac{2(1-\cos\theta)}{a},
\end{equation}
where $\theta$ is the scattering angle, $a=m_{e0}/k_1$, $m_{e0}$ is
the rest mass of the electron.

Now, consider the change of the wavelength of the photon:

Using Eq.(10), the de Broglie wavelength can be expressed as
\begin{equation}
\lambda=\frac{h}{2ck_1}\cdot \frac{1}{\sqrt{(a+n)n}}.
\end{equation}

For a photon, $a=0$, we have
\begin{equation}
\lambda=\frac{h}{2ck_1}\cdot \frac{1}{n}.
\end{equation}

So the Compton shift $\triangle \lambda$ is
\begin{equation}
\triangle
\lambda=\lambda'-\lambda=\frac{h}{2ck_1}(\frac{1}{n'}-\frac{1}{n}),
\end{equation}
where $\lambda$ and $\lambda'$ are the wavelengths of the incident
photon and of the scattered photon respectively.

Associating Eqs.(26), (29) and the relation $a=m_{e0}/k_1$ gives
\begin{equation}
\triangle \lambda=\frac{h}{m_{e0} c}(1-\cos\theta).
\end{equation}

\subsection{Creation and annihilation of electron-positron
pairs}

The idea of vacufluctons will provide us a simple description of the
creation and annihilation of electron-positron pairs. To illustrate
this, we will discuss the annihilation of an electron-positron pair
below. For simplicity, here we assume that the electron and the
positron are both at rest initially. The conservation of matter and
momentum gives
\begin{figure}
\centering
\includegraphics[totalheight=1.1in]{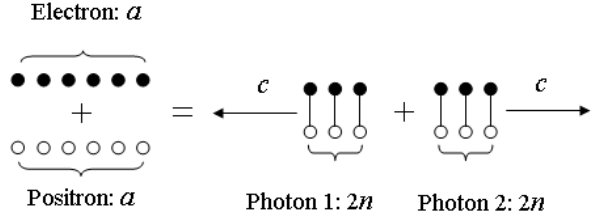}
\caption{\label{fig06} The annihilation of an electron-positron
pair.}
\end{figure}
\begin{eqnarray}
a+a=2n_1+2n_2, \nonumber \\
0=\vec{n}_1+\vec{n}_2.
\end{eqnarray}
We get
\begin{eqnarray}
\vec{n}_1=-\vec{n}_2, \nonumber \\
2n_1=2n_2=a.
\end{eqnarray}

That is, the electron-positron pair will change symmetrically into a
pair of photons moving towards two opposite directions. Fig.6 shows
a simplified description of this annihilation.

\section{Further deductions of the special theory of relativity}
\label{sec:4}

Compared with the usual method, the calculation of the Compton shift
and the electron-positron annihilation here is simple, which needs
no use of the relations $E=mc^2$ and $m=m_0/\sqrt
{1-\upsilon^2/c^2}$. In fact, these and some other relations given
by the special theory of relativity have already been embodied in
the postulates of vacuum fluctuation. To illustrate this, the paper
below will give the further deductions of the basic conclusions of
the special theory of relativity, based only on the three postulates
and their deductions given in Section 3.

\subsection{Mass-velocity relation}

Eqs. (7) and (9) give the velocity of a matter particle
\begin{equation}
\upsilon=2c \frac{\sqrt{(a+n)n}}{a+2n},
\end{equation}
where $a$ vacufluctons form the rest mass of a moving particle
\begin{equation}
m_0=k_1 \cdot a,
\end{equation}
and $2n$ vacufluctons form the increment of mass
\begin{equation}
\triangle m=k_1 \cdot 2n.
\end{equation}
The total mass is
\begin{equation}
m=k_1 \cdot (a+2n)=k_1 \cdot Q.
\end{equation}

Combining Eqs. (33), (34) and (36), we obtain the mass-velocity
relation
\begin{equation}
m=\frac{m_0}{\sqrt {1-\frac{\upsilon^2}{c^2}}}.
\end{equation}

\subsection{Mass-energy relation}

Substituting Eqs. (6), (3) and (17) into Eq. (18) gives the
mass-energy relation
\begin{equation}
E=mc^2.
\end{equation}

In fact, this relation can be directly deduced from the idea of
vacufluctons, as shown in Eqs. (10) to (15).

If $n=0$, we have the rest energy
\begin{equation}
E_0=m_0 c^2.
\end{equation}

Through Eqs. (2), (3) and (38) we get
\begin{equation}
d\sum E_i=0.
\end{equation}

It is just the conservation law of energy. In fact, the energy of
matter is a sum of fluctuation energy of all the vacufluctons. And
in average, there is no difference between the fluctuation energy of
each vacuflucton. Therefore, the energy of matter must be directly
proportional to its total number of vacufluctons. Thus the
conservation of vacufluctons of an isolated system certainly leads
to the conservation of energy.

\subsection{Energy-momentum relation}

Substituting Eqs. (36), (34) into Eqs. (38), (39) gives
\begin{equation}
E=k_1 (a+2n) c^2,
\end{equation}
and
\begin{equation}
E_0=k_1 a c^2.
\end{equation}

Then we have
\begin{equation}
E^2-E^2_0=[2ck_1 \sqrt{(a+n)n}]^2 c^2.
\end{equation}

Combining this with Eq. (10) gives the energy-momentum relation
\begin{equation}
E^2=p^2c^2+E^2_0.
\end{equation}

\subsection{Time dilation}

Eqs. (18) and (22) give the phase velocity of a matter wave
\begin{equation}
V_\varphi=\lambda \nu=\frac{h}{p}\cdot
\frac{E}{h}=\frac{E}{p}=\frac{mc^2}{m V_g},
\end{equation}
where $V_g$ is the group velocity of the vacufluctons, i.e., the
velocity $\upsilon$ of the matter particle.

Then we have
\begin{equation}
V_\varphi V_g=c^2.
\end{equation}

Generally we have $V_g \leqslant c$, so there is $V_\varphi
\geqslant c$.

The difference between the phase velocity and group velocity
indicates that there is a phase shift within the group of
vacufluctons. Assuming that this internal shift has a frequency of
$\nu_{in}$, we have
\begin{equation}
\nu_{in}=\frac{V_\varphi - V_g}{\lambda} =\nu \frac{V_\varphi -
V_g}{\lambda \nu}=\nu (1-\frac{V_g}{V_\varphi}).
\end{equation}

Associating Eqs. (33), (46) and (47) gives
\begin{eqnarray}
\nu_{in}&=&\nu \cdot \ [(\frac{a+2n}{2})^2-(a+n)n]/(\frac{a+2n}{2})^2\nonumber\\
&=&\nu \cdot \ (\frac{a}{2})^2/(\frac{a+2n}{2})^2.
\end{eqnarray}

The physical meaning of the above equation is clear. In the
equation, $\nu$ is the fluctuation frequency of the whole group of
vacufluctons, and $[(a+2n)/2]^2$ is the number of possible
positive-negative pairing in the ideal case as that of a photon
(that is, the number of positive vacufluctons is just equal to that
of negative vacufluctons, and the number of possible pairing reaches
a maximum), while the actual number of possible pairing is $(a+n)n$,
and $(a/2)^2$ is the difference between these two. Obviously, in
order to get the whole group of vacufluctons moving ahead at the
same velocity, it is needed to change the positive-negative
partnership between each other. It will therefore cause a phase
shift within the group. That is, there will be an internal frequency
of $\nu_{in}$, which can be a natural clock of the matter.

Substituting Eqs. (1) and (6) into Eq. (48) yields
\begin{equation}
\nu_{in}=k_3 a \cdot \frac{a}{a+2n}.
\end{equation}

The rest internal frequency of this vacuflucton group (i.e., $n=0$)
is
\begin{equation}
\nu_{in0}=k_3 a.
\end{equation}

Then we have
\begin{equation}
\nu_{in}=\nu_{in0}\cdot \frac{a}{a+2n}=\nu_{in0}\cdot
\frac{m_0}{m}=\nu_{in0}\sqrt {1-\frac{\upsilon^2}{c^2}}.
\end{equation}

It indicates that a moving clock runs slower than an identical
stationary clock. Considering the relation  $T=1/\nu$, where $T$
denotes the periodic time, then we have $T=T_0 /\sqrt
{1-\frac{\upsilon^2}{c^2}}$, which indicates that there is a time
dilation effect
\begin{equation}
\bigtriangleup t=\bigtriangleup t_0 /\sqrt
{1-\frac{\upsilon^2}{c^2}}.
\end{equation}

This effect can also be deduced through the Lorentz transformation.
But here, through the investigation of the internal periodicity of
the vacuflucton group, we get a more comprehensible view of the time
dilation effect.

\subsection{Length contraction}

According to Eq. (33), a photon ($a=0$) moves in vacuum at a
constant velocity of $c$, no matter which frame of reference is
selected. So we can choose a photon as a uniform tool to measure the
length of an object in different reference frames.
\begin{figure}
\centering
\includegraphics[totalheight=1.0in]{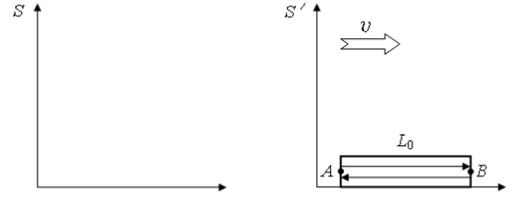}
\caption{\label{fig07} Measuring the length of an object with a
photon.}
\end{figure}

In Fig.\ 7, a rod $AB$ is at rest in the frame $S'$, which is moving
at velocity $\upsilon$ relative to the frame $S$. A photon moves
from $A$ to $B$ and returns to $A$. The time interval of this event
measured in $S'$ is
\begin{equation}
\bigtriangleup t_0=\frac{2L_0}{c},
\end{equation}
where $L_0$ is the length of the rod measured in $S'$.

Since points $A$ and $B$ are both moving at velocity $\upsilon$
relative to $S$, so when observed in frame $S$, the photon moves a
length of $L+\upsilon \bigtriangleup t_{AB}=c\bigtriangleup t_{AB}$
from $A$ to $B$, and a length of $L-\upsilon \bigtriangleup
t_{BA}=c\bigtriangleup t_{BA}$ back to $A$. Thus we get
\begin{eqnarray}
\bigtriangleup t_{AB}=\frac{L}{c-\upsilon}, \nonumber \\
\bigtriangleup t_{BA}=\frac{L}{c+\upsilon},
\end{eqnarray}
where $L$ is the length of the rod measured in $S$.

Therefore, the time interval of the event measured in $S$ is
\begin{equation}
\bigtriangleup t=\bigtriangleup t_{AB}+\bigtriangleup
t_{BA}=\frac{2L}{(c^2-\upsilon^2)/c}.
\end{equation}

Through Eqs. (53) and (55) we get
\begin{equation}
\frac{\bigtriangleup t}{\bigtriangleup t_0}=\frac{L}{L_0} \cdot
\frac{1}{1-\upsilon^2/c^2}.
\end{equation}

Substituting Eq. (52) into the above equation gives
\begin{equation}
L=L_0 \sqrt{1-\frac{\upsilon^2}{c^2}}.
\end{equation}

It is just the length contraction of a moving object.

\section{Testing the postulates and determining the constants}
\label{sec:5}

Eq.(28) gives the wavelength of a photon as: $\lambda=h/2nck_1$.
Here we know that the product of the number of positive-negative
paired vacufluctons of a photon and the wavelength of the photon is
a constant, i.e., $2n\lambda=h/ck_1$. Usually, the number $2n$ is
extremely large, so the wavelengths can almost form a continuous
spectrum. However, under the condition of extremely low frequency,
the discreteness of the electromagnetic spectrum may be observed
ultimately. In such a case, every two wavelengths will be a ratio of
two not-too-large integers, that is, $\lambda_1/\lambda_2=n_2/n_1$.
If such a relation is discovered in experiments, the postulates of
vacufluctons can then be proved to be true.

Also, through the analysis of the data collected in such
experiments, we will finally find the definite relation between
$\lambda$ and $n$. Then we could find out the value of constant
$k_1$, and figure out the constants $k_2$ and $k_3$ through the
relations $k_2/2k_1=c$ and $c^2k_1/k_3=h$.

Considering that the frequency of the lowest electromagnetic waves
people have already detected is on the order of the $10^{-3}$ Hz
\cite{Elert2001}, as an upper estimate, we have $k_1=h/2n\lambda
c=h\nu/2nc^2<h\nu/2c^2=3.68 \times 10^{-54} \textnormal{kg}$. It is
already an extremely small mass --- a mass 23 orders smaller than
that of the electron.

\section{Conclusions}
\label{sec:6}

The three postulates of vacuum fluctuation proposed in this paper
provide an insight into the relationship between matter and vacuum.
Vacuum is not just a place where particles stay and move; as a
matter of fact, vacuum is the origin of all matter, motion and
evolution. Therefore, the photon, electron and other matter
particles could be unified at the level of vacuum fluctuational
particles, or as we called, vacufluctons.

In Einstein's theory of relativity, it is just an assumption that
the velocity of light in vacuum is a constant and is the limit of
all the moving particles. While in this paper, it can be deduced
naturally from the postulates of vacuum fluctuation.

In the view of vacuum fluctuation, the Planck constant can be
understood more profoundly. The value of the Planck constant is
determined by the constants $k_1$, $k_2$ and $k_3$, that is,
$h=k_2^2/4k_1k_3$. And the essential of the Planck constant is that
it characterizes the generality and unity of vacuum fluctuation.
Relatedly, the randomicity and periodicity of the fluctuation of
vacufluctons form the intrinsic characteristics of matter, leading
to the so puzzling properties of uncertainty and wave-particle
duality. The convenient deduction of the energy-frequency relation
and the de Broglie wavelength expression from the postulates of
vacufluctons shows the rationality of the ideas.

The simple calculation of the Compton shift and the
electron-positron annihilation also shows the power of the three
postulates. To illustrate this simplicity, the basic conclusions of
the special theory of relativity such as the relations of
mass-velocity, mass-energy, energy-momentum, time dilation and
length contraction are further deduced. It is not that strange to
see all the results being fully in agreement with that obtained
through the Einstein's theory of relativity, because the basic
premise of the special theory of relativity, that is, the constant
velocity of light in vacuum, is just one of the deductions of the
postulates of vacuum fluctuation.

In conclusion, it is suggested that the same physical basis of the
theory of relativity and the quantum mechanics will be found through
the investigation of vacuum fluctuation. Actually, a great many of
results can be naturally deduced from the basic postulates of vacuum
fluctuation. Furthermore, the investigation of vacuum fluctuation
will open up a new way to deal with the inexplicable problems in
physics today. In brief, it promises an approach to the unification
of physical theories.



\begin{acknowledgments}
I wish to acknowledge the supports from the Ministry of Science and
Technology of China (grant no. 2006CB921403 \& 2006 AA06A204) and
the Zhejiang Provincial Qian-Jiang-Ren-Cai Project of China (grant
no. 2006R10025).
\end{acknowledgments}

\end{document}